\documentclass[12pt, final]{asme2ej}

\usepackage{epsfig} 
\usepackage{graphicx}
\usepackage{amssymb}
\usepackage{subfigure}
\usepackage{color}
\usepackage{algorithm}

\title{Inflow turbulence generation for eddy-resolving simulations of turbomachinery flows}

\author{Sunil K. Arolla
    \affiliation{
	Postdoctoral research associate\\
	Sibley School of Mechanical and Aerospace Engineering\\
	Cornell University\\
	Ithaca, NY 14853\\
    Email: ska62@cornell.edu
    }	
}

\begin{document}

\maketitle    

\begin{abstract}
{\it 
A simple variant of recycling and rescaling method to generate inflow turbulence using unstructured grid CFD codes is presented. The method has been validated on large eddy simulation of spatially developing flat plate turbulent boundary layer. The proposed rescaling algorithm is based on the momentum thickness which is more robust and essentially obviates the need of finding the edge of the turbulent boundary layer in unstructured grid codes. Extension of this algorithm to hybrid RANS/LES type of approaches and for wall-bounded turbomachinery flows is also discussed. Results from annular diffuser with different inflow boundary layer characteristics is presented as an example application to show the utility of such an algorithm.
}
\end{abstract}

\section{Introduction}\label{sec:intro}
High-fidelity eddy-resolving simulations require specification of accurate and realistic inflow conditions. The inflow boundary layer thickness can have significant influence on the flow characteristics downstream. For example, in inter-turbine or inter-compressor diffuser configurations relevant to turbomachinery, the inlet boundary layer thickness determines the flow behavior through the diffuser \cite{stevens:1980}. Hence, addressing this issue with a robust approach that can be used within general unstructured CFD codes is critical.

Numerical simulations of fully developed, time-evolving flows are often performed using periodic boundary conditions in which the downstream flow can be directly re-applied at the inlet. However, these boundary conditions are not appropriate for spatially developing flows, such as turbulent boundary layers. In simulating such flows, the flow downstream is highly dependent on the conditions at the inlet, making it necessary to specify a realistic time series of turbulent fluctuations that are in equilibrium with the mean flow. The inflow data should satisfy the Navier-Stokes equations to be accurate.

The most straightforward approach to simulate a spatially developing turbulent boundary layer is to start the calculation far upstream with a laminar profile plus random disturbances and then allow for \emph{natural transition} to turbulence to occur. This method is not generally applicable for turbulent flow simulations as it requires a long development section to simulate \emph{natural transition} and hence is prohibitively expensive. The other simple procedure for specifying turbulent inflow conditions is to superimpose random fluctuations on a desired mean velocity profile. The amplitude of the turbulent fluctuations can be adjusted to satisfy a desired set of one-point second order statistics. However, the velocity derivative skewness is zero and hence inflow condition is void of nonlinear energy transfer and the flow lacks realistic turbulent structure. Also, a fairly lengthy development section is required to allow for development of organized turbulent motion. In addition, it is often hard to control the skin friction and integral boundary layer thickness a the end of the development section.

The method of using an auxiliary simulation to generate inflow boundary conditions is commonly used for internal flows \cite{akselvoll:1995}. A similar approach can be used for turbulent boundary layers as well. To account for spatial growth, Spalart (1988) developed a method by adding source terms to the Navier-Stokes equations \cite{spalart:1988}. This method is capable of producing equilibrium turbulent boundary layer with direct control on skin friction and integral boundary layer thickness. However, it requires a coordinate transformation that minimizes the streamwise inhomogeneity and hence cannot be adopted into general purpose CFD codes.

Lund \emph{et al.} (1998) proposed the widely used \emph{recycling and rescaling method} in which the velocity at the inflow plane is estimated using the flow downstream \cite{lund:1998}. The velocity field extracted at a downstream location is rescaled and reintroduced at the inlet. This method proved to be very successful in generating accurate inflow data with specific boundary layer thickness. Some of the numerical issues reported in the literature with the Lund \emph{et al.} method are: spurious spanwise structures are recycled that can grow in time and disrupt the numerical stability, sensitivity to the initialization. Several different strategies have been adopted to address these issues such as using dynamically shifting the recycling location \cite{liu:2006}, constant spanwise shift \cite{spalart:2006}, constant spanwise reflection \cite{jewkes:2011}, dynamic shifting and reflection using a random-walk method \cite{morgan:2011}. Ferrante and Elghobashi (2004) presented a modified method by imposing a specific energy spectrum to insure the statistical correlation between the streamwise and wall-normal fluctuations a non-vanishing magnitude \cite{ferrante:2004}.

Synthetic methods form another class of generating inflow conditions. These methods are characterized by the use of some model to prescribe turbulent fluctuations about a mean flow profile. Yao and Sandham (2002) proposed one of the first synthetic methods in which the observed features in a turbulent boundary layer such as near-wall and lifted streaks are semi-analytically prescribed by enforcing perturbation velocities according to the superposition of several waveform functions \cite{yao:2002}. These waveform modes have amplitudes and phase shifts that correspond to desired streak lengths and thicknesses. Klein \emph{et al.} have developed a digital filtering approach which is also widely used \cite{klein:2003}. These methods however require sufficiently long domain lengths for the turbulent flow to recover from modeling errors. In addition, \emph{a priori} knowledge of mean flow, Reynolds stresses is required for using these methods. Additional synthetic methods exist that offer shorter recovery lengths \cite{jarrin:2006, pamies:2009}. A more detailed review of different inflow generation methods can be found in \cite{keating:2004}. This article is concerned with \emph{recycling and rescaling} type of approach in which \emph{a priori} knowledge of mean flow and turbulence statistics is not required.

Spalart \emph{et al.} (2006) proposed a variant of \emph{recycling and rescaling} method where several physical arguments have been used to simplify the algorithm \cite{spalart:2006}. Lund \emph{et al.} (1998) method uses different scaling laws for inner and outer layers. It also involves decomposition of the velocity field into mean and fluctuating components. Spalart \emph{et al.} argue that the near-wall turbulence regenerates itself much faster than the outer region and hence proposed to use outer layer scaling throughout. Also, the rescaling is applied only the streamwise velocity components as corrections to the wall-normal components have very little effect. The advantage of this method is the spatially developing simulation generates its own inflow conditions and a short recycling distance leads to a reduction in computational cost. This method has been used in investigating Spalart-Allmaras model based detached eddy simulation (DES) models with ambiguous grid densities \cite{spalart:2006a}. A comparison between \cite{spalart:2006} and \cite{lund:1998} methods is presented in \cite{arolla:2014}.

Use of such inflow generation methods with unstructured grid CFD codes in the context of turbomachinery applications has not been reported in the literature. In this article, a simple variant of recycling and rescaling method for generating inflow turbulence for unstructured grid CFD codes is presented and validation on large eddy simulation (LES) of flat plate turbulent boundary layer is reported. Extension of the method to hybrid RANS/LES type of approaches is discussed. As an example for such an approach, a recently proposed simplified version of Improved Delayed Detached Eddy Simulation (IDDES) of \cite{gritskevich:2012} is implemented and applied for spatially developing turbulent boundary layer using the proposed inflow generation method. For applying this method for turbomachinery applications, the required modifications are presented and validated on annular diffuser problem.

\section{Computational framework}
The computational framework used in this research is that of OpenFOAM finite volume based incompressible flow solver. The filtered Navier-Stokes equations solved in the context of LES are:

\begin{eqnarray}
\partial_{i}\hat{u_{i}} &=& 0 \\
\partial_{t}\hat{u_{i}}+\partial_{j}\hat{u_{j}}\hat{u_{i}} &=& -\frac{1}{\rho}\partial_{i}\hat{p} + \nu \nabla^{2}\hat{u_{i}} - \partial_{i}\tau_{ij}^{SGS}
\end{eqnarray}

where $\hat{u_{i}}$ is the filtered velocity field. The unclosed term that arises due to filtering operation are the subgrid scale stresses given by $\tau_{ij}^{SGS}$. The equations are close by employing a dynamic Smagorinsky model \cite{germano:1991} with modification by Lilly (1992) \cite{lilly:1992}.

As an example for hybrid RANS/LES approaches, a recently proposed simplified version of IDDES for $k-\omega$ SST model \cite{gritskevich:2012} has been implemented within OpenFOAM framework. Stated briefly, the transport equations for turbulent kinetic energy ($k$) and specific dissipation rate ($\omega$) are of the following form:

\begin{eqnarray}
\frac{\partial k}{\partial t} + u_{j} \frac{\partial k}{\partial x_{j}} = P_{k}- \sqrt{k^{3}}/l_{IDDES} + \frac{\partial}{\partial x_{j}}\left[\left(\nu+\frac{\nu_{T}}{\sigma_{k}} \right) \frac{\partial k}{\partial x_{j}} \right] \\
\newline
\frac{\partial \omega}{\partial t} + u_{j} \frac{\partial \omega}{\partial x_{j}} = P_{\omega}- D_{\omega} + CD_{\omega} + \frac{\partial}{\partial x_{j}}\left[\left(\nu+\frac{\nu_{T}}{\sigma_{\omega}} \right) \frac{\partial \omega}{\partial x_{j}} \right]
\end{eqnarray}
where $P_{k}$ is the production of turbulent kinetic energy, $P_{\omega}$ is the production of specific dissipation rate, $D_{\omega}$ is the dissipation of $\omega$, and $CD_{\omega}$ is the cross-diffusion term in $\omega$. The term $l_{IDDES}$ is the length scale that operates the switch between RANS and LES. The eddy viscosity is of the form $\nu_{T}={k}/{\omega}$ with a limiter for separated flows. For detailed description of these terms and empirical constants, see \cite{gritskevich:2012}. The governing equations solved are similar to that of LES, but subgrid scale stress term is replaced by a modeled Reynolds stress.

For the numerical simulations presented in this article, Pressure Implicit with Splitting of Operator (PISO) algorithm is employed. A second order accurate backward implicit scheme for time discretization and a second order central scheme (with filtering for high-frequency ringing) for spatial discretization is used. The initial and boundary conditions are discussed for each validation problem in the subsequent sections.

\section{A variant of recycling and rescaling method for inflow turbulence generation}
The recycling and rescaling method by \cite{lund:1998} uses scaling laws by dividing the boundary layer into inner and outer regions. A composite profile is derived using a weighting function based on hyperbolic tangent function. The scaling operation requires computation of friction velocity and to obtain required momentum thickness, one must iteratively adjust the boundary layer thickness until the target value is reached.

The essential idea presented in this paper is based on the work by \cite{spalart:2006} to simplify the inflow generation algorithm based on the following physical arguments:
\begin{itemize}
\item The near-wall turbulence regenerates itself much faster than the outer region turbulence $\to$ Apply outer layer scaling throughout.
\item When the recycling station is located quite close to the inflow, which is desirable in terms of computing cost, the conflict between inner and outer region scaling essentially vanishes $\to$ Short recycling distance
\item Corrections to the wall normal velocity component $v$ have very little effect $\to$ Omitted
\end{itemize}

\begin{figure}[!htp]
\centering{
\includegraphics[width=80mm]{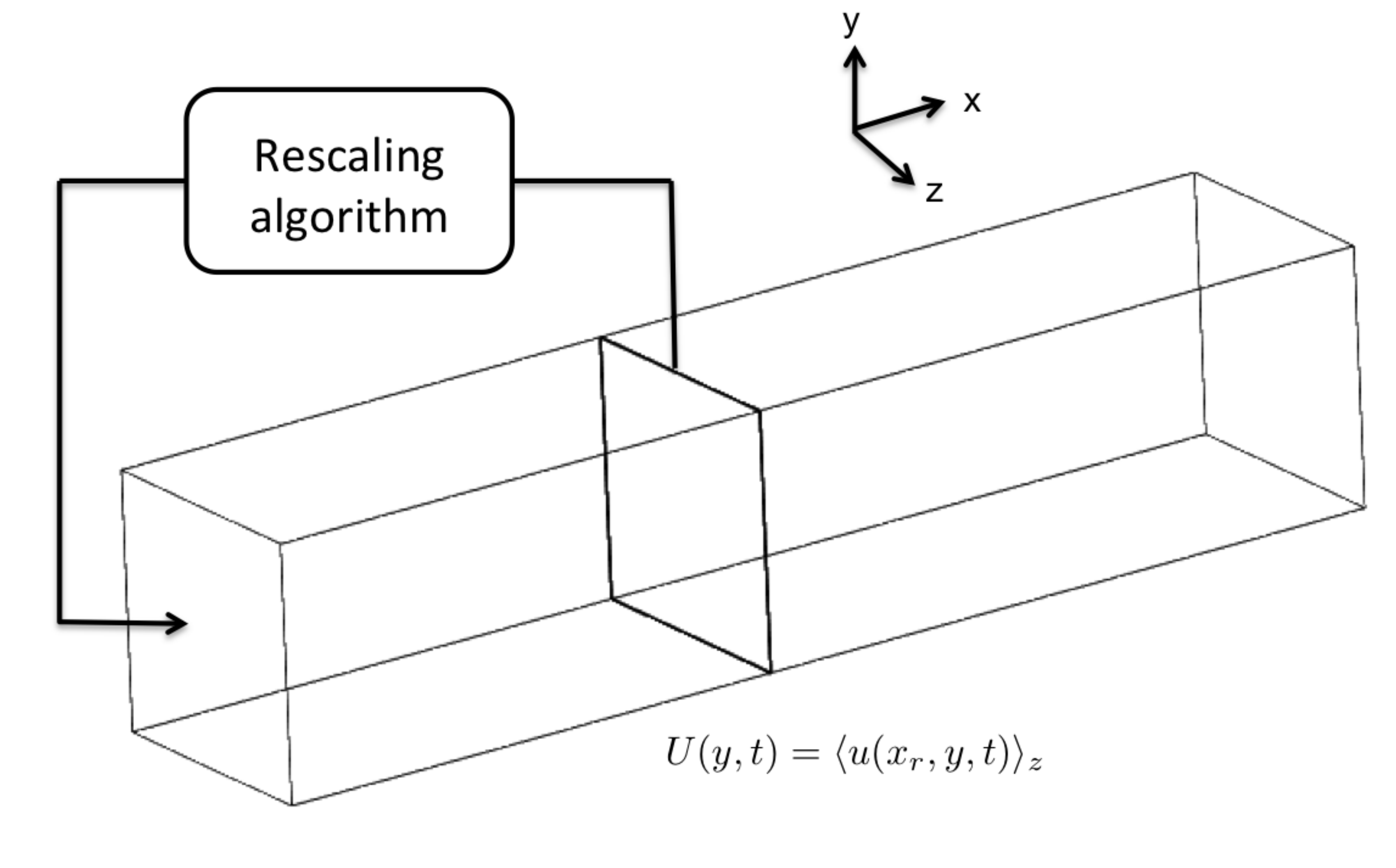}
}
\caption{A schematic of the computational domain used for flat plate turbulent boundary layer simulation. The recycling plane is located at $x_{r}=5 \delta_{0}$ from the inflow boundary.}
\label{flatplate_config}
\end{figure}

In the current work, momentum thickness based scaling is used in place of $99\%$ boundary layer thickness. This avoids the need of locating the edge of the boundary layer thickness. Moreover, using integral quantities like momentum thickness (or displacement thickness) is numerically robust. Most experiments report the momentum thickness Reynolds number at the inflow and hence back-to-back simulations can be set-up easily. A spanwise mirroring method \cite{jewkes:2011} is adopted as it was found to be adequate in the current work for disorganizing unphysical spanwise durable structures. It should be noted that more advanced strategies like random-walk based dynamic shifting and reflection might be more efficient, but those have not been tried out in the present work.

The steps involved in the inflow generation algorithm are:
\begin{enumerate}
\item Extract the velocity field, $u(x_{r},y,z,t)$, at the recycling plane located at $x_{r}$ and project on to the inflow boundary (see figure \ref{flatplate_config}).
\item Perform spanwise averaging to get $U(y,t) = \langle u(x,y,t) \rangle_{z}$. A simple indexing algorithm is used for the averaging. It involves looping over all the faces and index faces with the same wall-normal coordinate. Since the recycling plane is fixed, this indexing can be stored at the preprocessing step itself and reused at each timestep.
\item Find the freestream velocity $U_{\infty}=U(y_{max},t)$.
\item Integrate the velocity profile to compute the momentum thickness:
\begin{equation}
  \theta_{r} = \int_{0}^{y_{max}}\frac{U(y,t)}{U_{\infty}}\left(1-\frac{U(y,t)}{U_{\infty}} \right) dy
\end{equation}
\item Compute the rescaling factor, $\gamma = \theta_{r}/\theta_{in}$, where $\theta_{in}$ is the desired momentum thickness at the inflow.
\item Rescale only the x-component of the velocity field:
\begin{equation}
  u(x_{in},y,z,t) = u(x_{r},y\gamma,z,t-\Delta t)
\end{equation}
where $t-\Delta t$ means the velocity from the previous time step is used for convenience. A linear interpolation is used to compute velocity at the rescaled y-coordinate.

\item Apply spanwise mirroring to disorganize unphysical structures which would otherwise be recycled and take much time to get dampened by the spanwise diffusion.

\begin{eqnarray}
  u(x_{in},y,z,t) = u(x_{in},y,\Delta z-z,t)  \nonumber \\
  v(x_{in},y,z,t) = v(x_{in},y,\Delta z - z,t) \nonumber  \\
  w(x_{in},y,z,t) = -w(x_{in},y,\Delta z - z,t)
\end{eqnarray}
where $\Delta z$ is considered to be equal to the spanwise period. Note that $w$ has to be negative to ensure spatial coherence once mirrored \cite{jewkes:2011}.

\item Check for constant mass flow rate at the inflow by verifying the bulk velocity.
\end{enumerate}

The simplicity of this algorithm makes it amenable for extending to more complex applications as discussed further in this paper.

\subsection{Extended algorithm for hybrid RANS/LES type of approaches}
The algorithm presented in the previous section is mainly intended for direct numerical simulation (DNS) or large eddy simulation (LES). To extend this for hybrid RANS/LES type of approaches, the rescaling operation has to be modified depending on the the underlying RANS model.

For the Improved Delayed Detached Eddy Simulation (IDDES) model considered in this work, the rescaling operation on the underlying SST variant of $k-\omega$ model requires the following:

\begin{equation}
  k(x_{in},y,z,t) = k(x_{r},y\gamma,z,t-\Delta t)
\end{equation}

\begin{equation}
  \omega(x_{in},y,z,t) = \omega(x_{r},y\gamma,z,t-\Delta t)
\end{equation}
with $\gamma = \theta_{r}/\theta_{in}$.

\begin{figure}[!htp]
\centering{
\includegraphics[width=80mm]{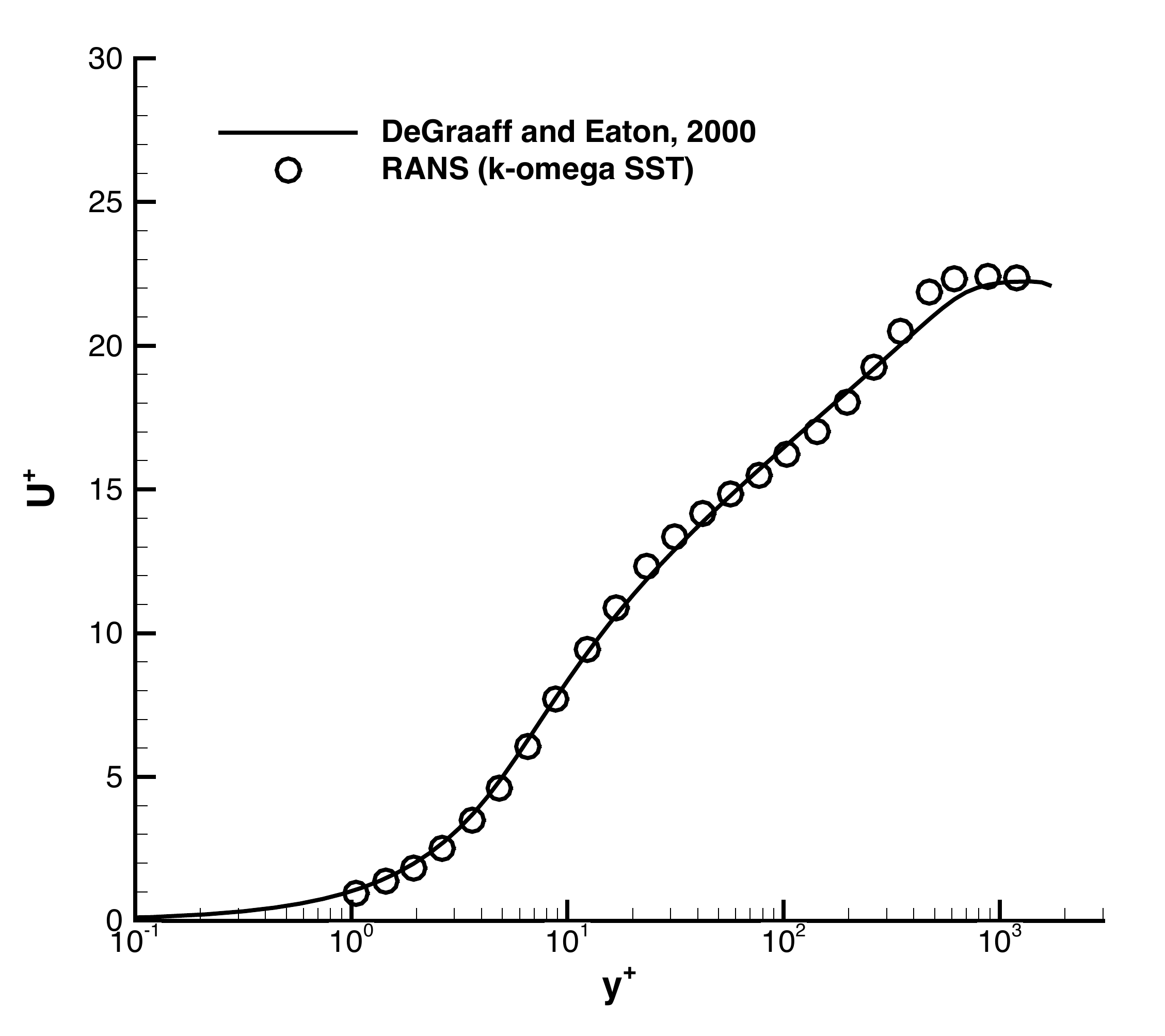}
}
\caption{RANS simulation of flat plate boundary layer using recycling and rescaling procedure: Mean velocity profile in wall units}
\label{Uplus_RANS}
\end{figure}

To verify the accuracy, this recycling and rescaling method has been applied for the RANS simulation of flat plate turbulent boundary layer. As shown in figure \ref{Uplus_RANS}, the method was able to generate a turbulent velocity profile accurately. So, no special care is taken to account for the location where the switching from RANS to LES takes place. The grid sensitivity of such hybrid RANS/LES approaches is still an open question and hence any errors associated with using hybrid methods for spatially developing boundary layers might be due to the underlying modeling assumptions and not the inflow generation method \emph{per se}.

The modifications required for using this algorithm for wall-bounded turbulent flows is discussed in the following subsection. Accuracy of this method for RANS approach means that, this could be used for imposing asymmetric velocity and turbulence profiles at the inlet. This is especially useful where the experimental data has inherent asymmetry due to the wind tunnel sidewall effects. This algorithm provides a means to set-up simulations consistent with the experiments and hence is useful in robust evaluation of the turbulence closure models used in the design.

\subsection{Extended algorithm for wall-bounded turbulent flows applied to turbomachinery}
Wall-bounded turbulent flows are often simulated using a streamwise periodic boundary condition to achieve fully developed turbulence condition at the inlet. But, in realistic turbomachinery applications such as inter-turbine or inter-compressor diffusers, it is important to specify a specific boundary layer thickness.

Since there is a variation in pressure in the streamwise direction, the validity of the scaling laws used in the inflow generation method becomes questionable. The momentum thickness based scaling used in the proposed algorithm can be extended to wall-bounded flows by making the following assumptions:
\begin{itemize}
\item Using radial coordinate instead of Y-coordinate.
\item Velocity at the half the height of annular diffuser is considered to be freestream velocity for computing momentum thickness.
\item Effect of transverse curvature is assumed negligible for the inflow generation purpose.
\item Effect of streamwise pressure gradient is ignored as a short recycling distance is used.
\end{itemize}

\begin{figure}[!htp]
\begin{minipage}[t!]{80mm}
\centering{
\subfigure[Schematic of the computational domain]{\includegraphics[width=80mm]{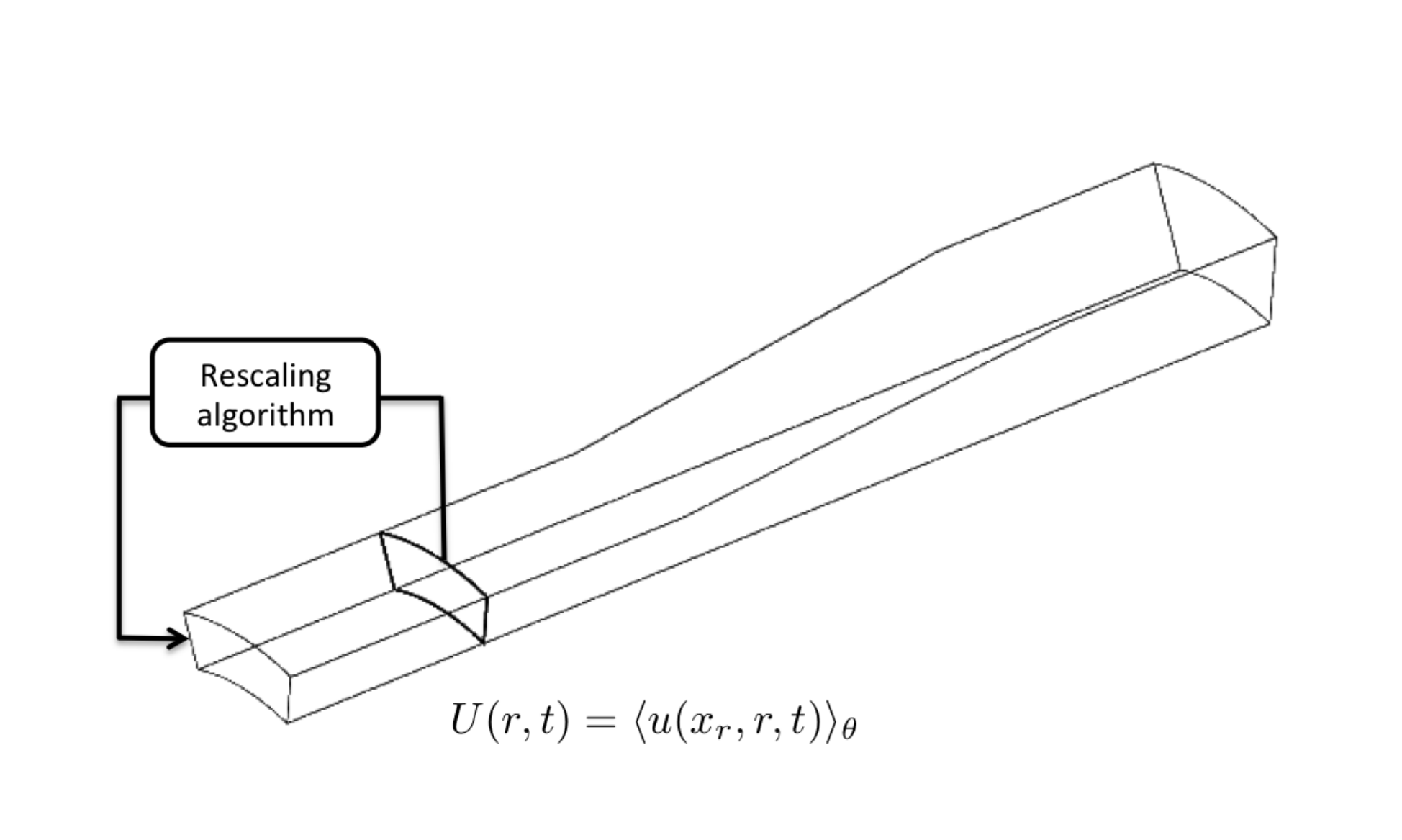}}
}
\end{minipage}
\hfill
\begin{minipage}[t!]{60mm}
\centering{
\subfigure[Nomenclature used in the algorithm]{\includegraphics[width=60mm]{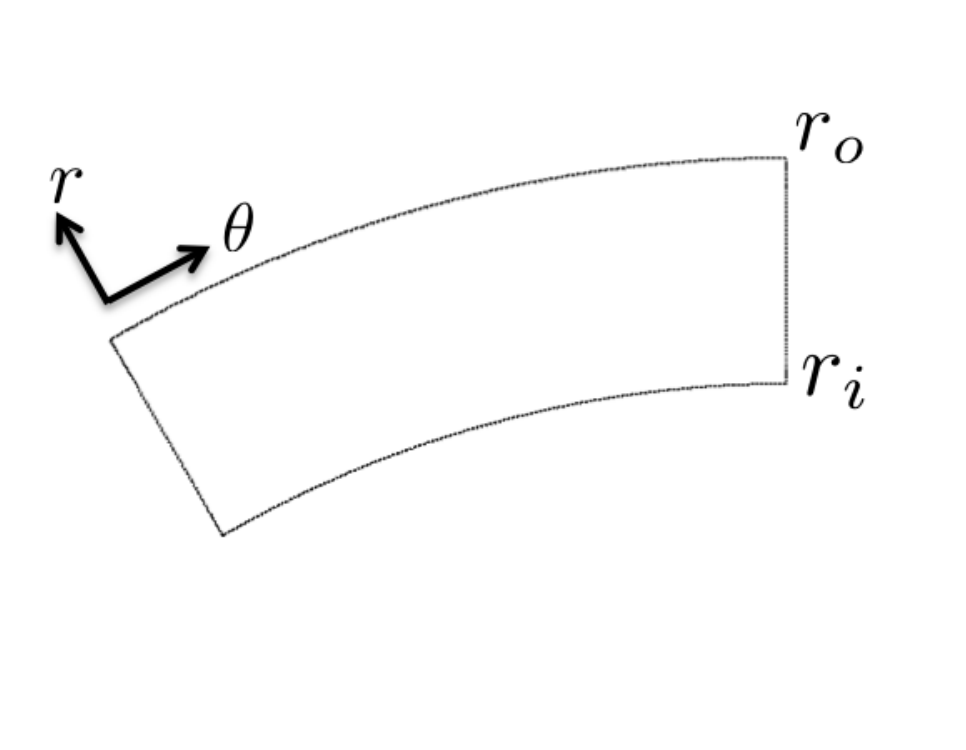}}
}
\end{minipage}
\caption{LES of $30^\circ$ sector of the annular diffuser}
\label{annulardiffuser_config}
\end{figure}

To adapt the inflow generation algorithm for annular diffuser type of applications, rescaling operation is applied separately for the hub and casing boundary layers. The momentum thickness for the hub and casing boundary layers is calculated as:

\begin{equation}
  \theta_{r} = \int_{0}^{r_{0.5}}\frac{U(r,t)}{U_{\infty}}\left(1-\frac{U(r,t)}{U_{\infty}} \right) dr
\end{equation}
where $U_{\infty}=U_{0.5}$ and $r_{0.5}=(r_{o}-r_{i})/2$. For the casing boundary layer, the velocity profile is integrated down to the half of the annulus height.

The accuracy of the algorithm for realistic turbomachinery applications is discussed in the following section.

\section{Results}
The proposed algorithm has been validated on eddy resolving simulations of flat plate turbulent boundary layer. As an example for a practical application, inlet conditions generated for LES of flow through annular diffuser are also presented and compared with the available experimental data. The mesh employed for these problems is structured, but is stored in an unstructured grid format for OpenFOAM, and hence the proposed algorithm is applicable for general unstructured CFD codes.

\subsection{Problem 1: Spatially developing flat plate turbulent boundary layer}

\subsubsection{LES results}

\begin{figure}[!htp]
\begin{minipage}[t!]{80mm}
\centering{
\subfigure[Mean velocity in wall units]{\includegraphics[width=80mm]{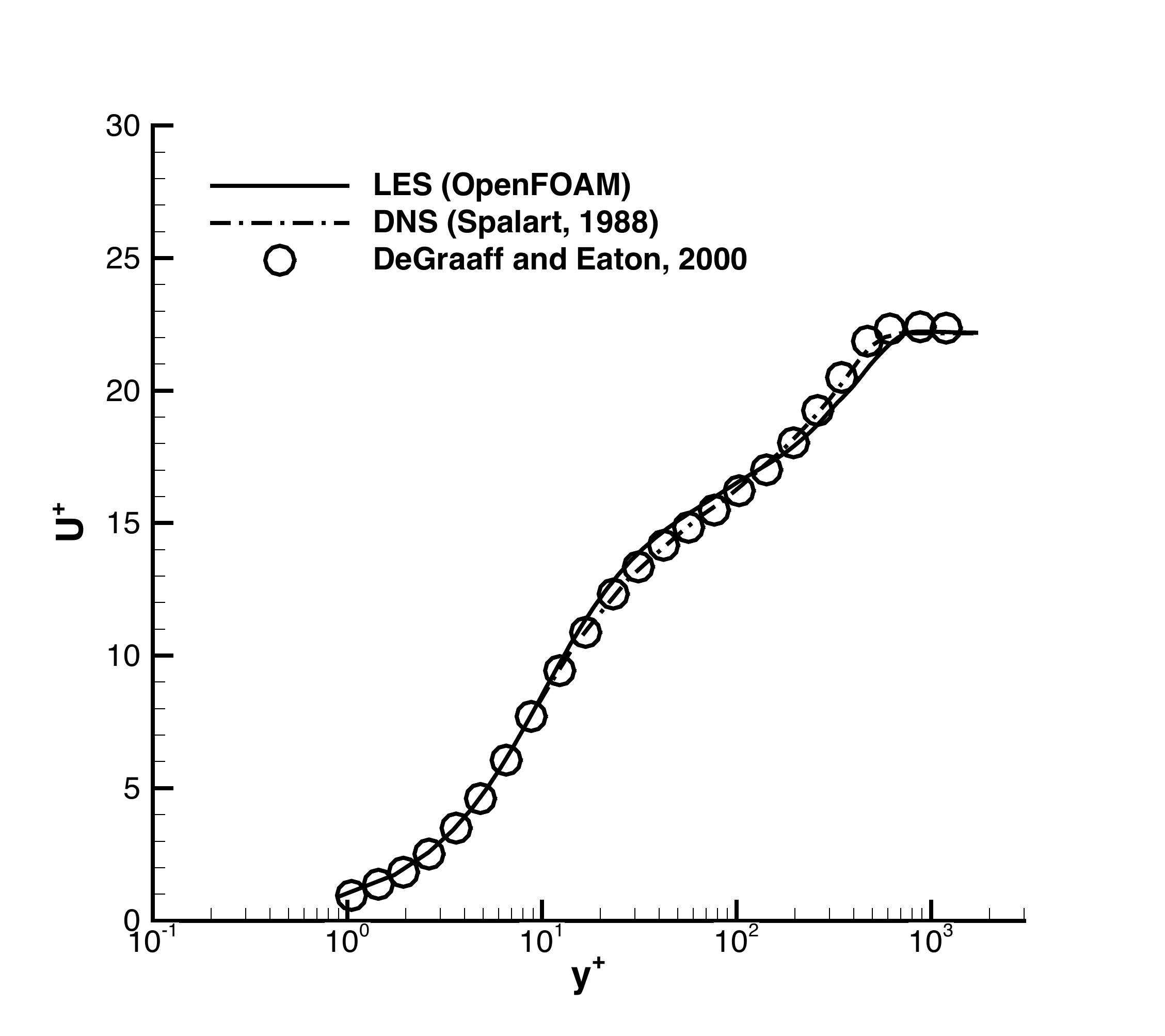}}
}
\end{minipage}
\hfill
\begin{minipage}[t!]{80mm}
\centering{
\subfigure[Reynolds stress components in wall units]{\includegraphics[width=80mm]{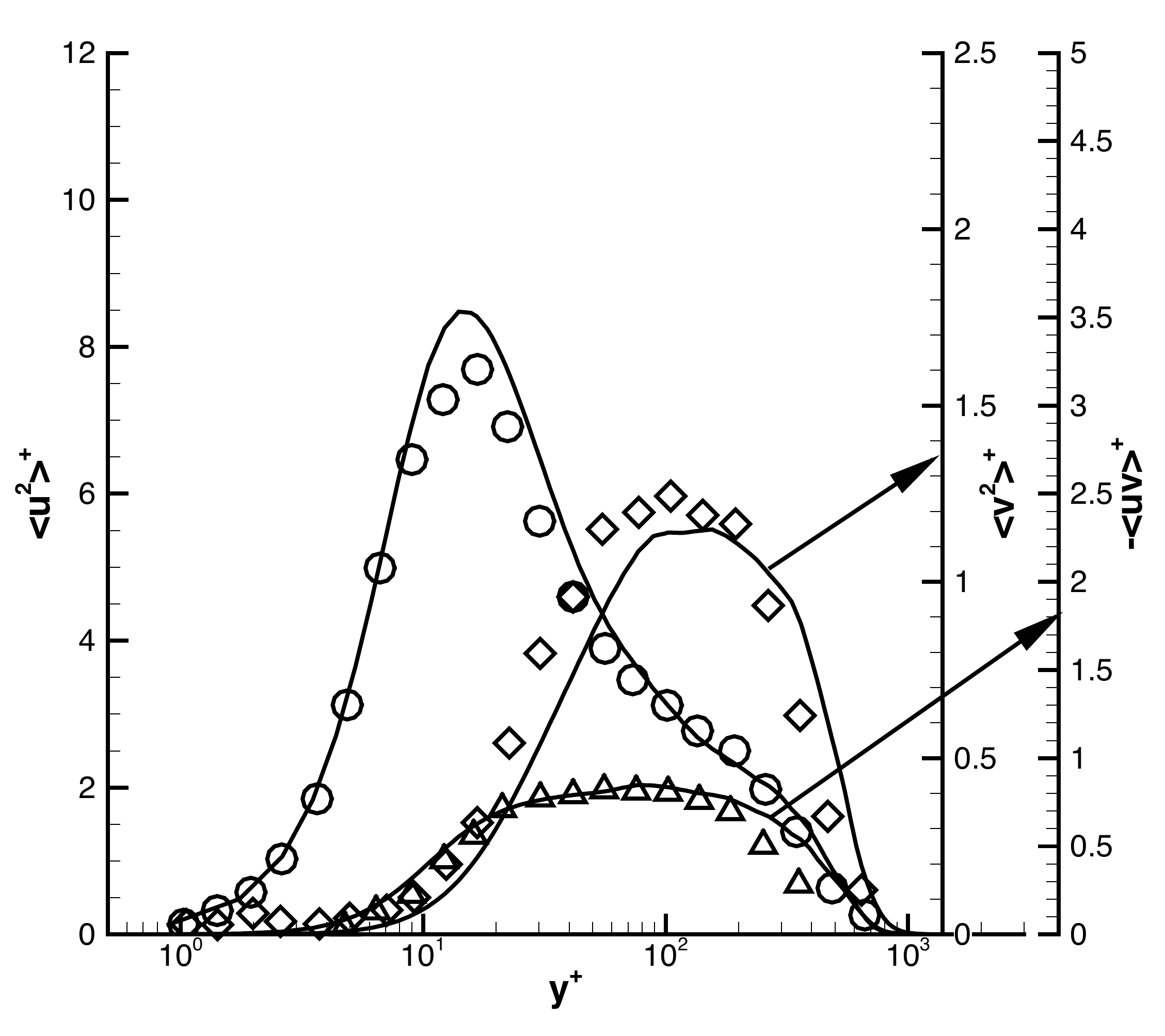}}
}
\end{minipage}

\caption{LES of spatially developing turbulent boundary layer: one-point statistics}
\label{LES_umean}
\end{figure}

As a baseline validation, results from LES of flat plate turbulent boundary layer with inflow momentum thickness Reynolds number of $R_{\theta}=1520$ are presented. The computational domain has dimensions $12\delta_{0} \times 3\delta_{0} \times 3\delta_{0}$ in the streamwise, wall-normal, and spanwise directions, respectively where $\delta_{0}$ is the $99 \%$ boundary layer thickness at the recycling plane. The mesh contains $182 \times 96 \times 164$ points in the streamwise, wall-normal, and spanwise directions, respectively. In terms of the wall units, the mesh resolution is $\Delta x ^{+} \approx 45 $, $\Delta y_{wall} ^{+} \approx 1$, $\Delta y_{max} ^{+} \approx 20$, and $\Delta z ^{+} \approx 12$. Uniform mesh is used in the streamwise and spanwise directions while a hyperbolic tangent stretching is used in the wall-normal direction to cluster points close to the wall. The recycling station was located at $5 \delta_{0}$ downstream of the inlet and the simulation provides its own inflow. The bottom wall is treated as a no-slip wall, top boundary is a slip wall, and at the outflow an advective boundary condition is used.

As noted in \cite{spalart:2006}, the initialization is important when using such inflow generation algorithms. The mean velocity profile given by Spalding law with random fluctuations with a maximum amplitude of $10\%$ of the freestream value superimposed on the mean value. The time step used is approximately two viscous time units ($\Delta t \approx 2\nu/u_{\tau}^{2}$). The simulation was run for 1000 inertial timescales ($\delta_{0}/U_{\infty}$) to eliminate transients and the statistics are collected over another 1000 timescales.

Figure \ref{LES_umean} presents comparison of the mean streamwise velocity and three Reynolds stresses plotted in wall units with the experimental data of \cite{degraaff:2000} for a flat plate boundary layer at $R_{\theta}=1430$. The mean velocity profile is in good agreement with the experimental profile as well as the DNS of \cite{spalart:1988}. The normal Reynolds stresses also show good agreement for the current grid resolution chosen. The shear stress shows much better agreement than that published in the earlier literature with LES.

\subsubsection{IDDES results}
The mesh used for IDDES for the same computational domain has $140 \times 96 \times 116$ points in the streamwise, wall-normal, and spanwise directions, respectively. In terms of wall units, the mesh resolution is $\Delta x ^{+} \approx 60 $, $\Delta y_{wall} ^{+} \approx 1$, $\Delta y_{max} ^{+} \approx 20$, and $\Delta z ^{+} \approx 16$. It was found that the recycling plane needs to be much closer to the inflow boundary for IDDES to sustain turbulence. In the current work, $2-3 \delta_{0}$ was found to be optimal recycling distance for IDDES.

\begin{figure}[!htp]
\begin{minipage}[t!]{80mm}
\centering{
\subfigure[Mean velocity in wall units]{\includegraphics[width=80mm]{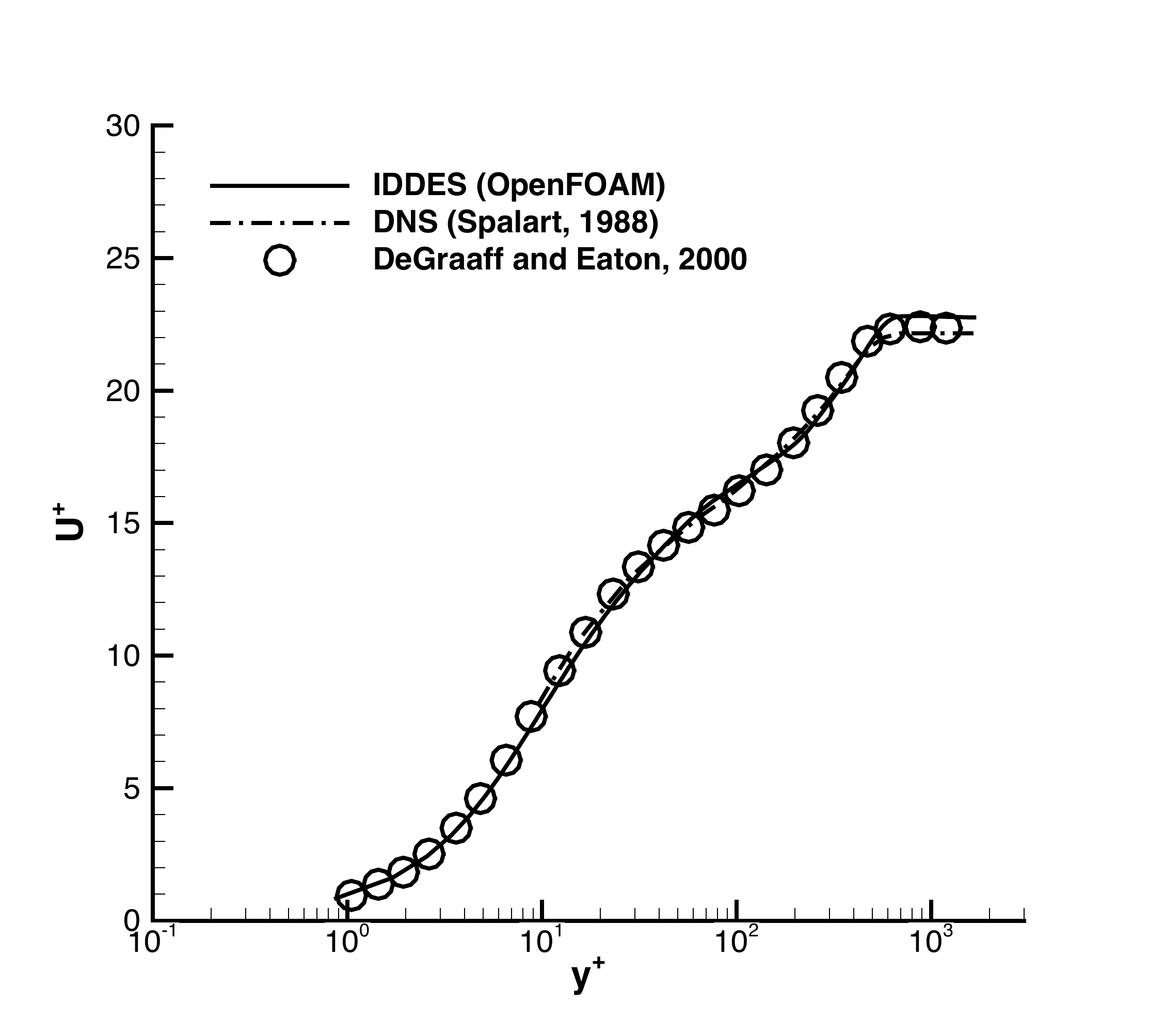}}
}
\end{minipage}
\hfill
\begin{minipage}[t!]{80mm}
\centering{
\subfigure[Reynolds stress components in wall units]{\includegraphics[width=80mm]{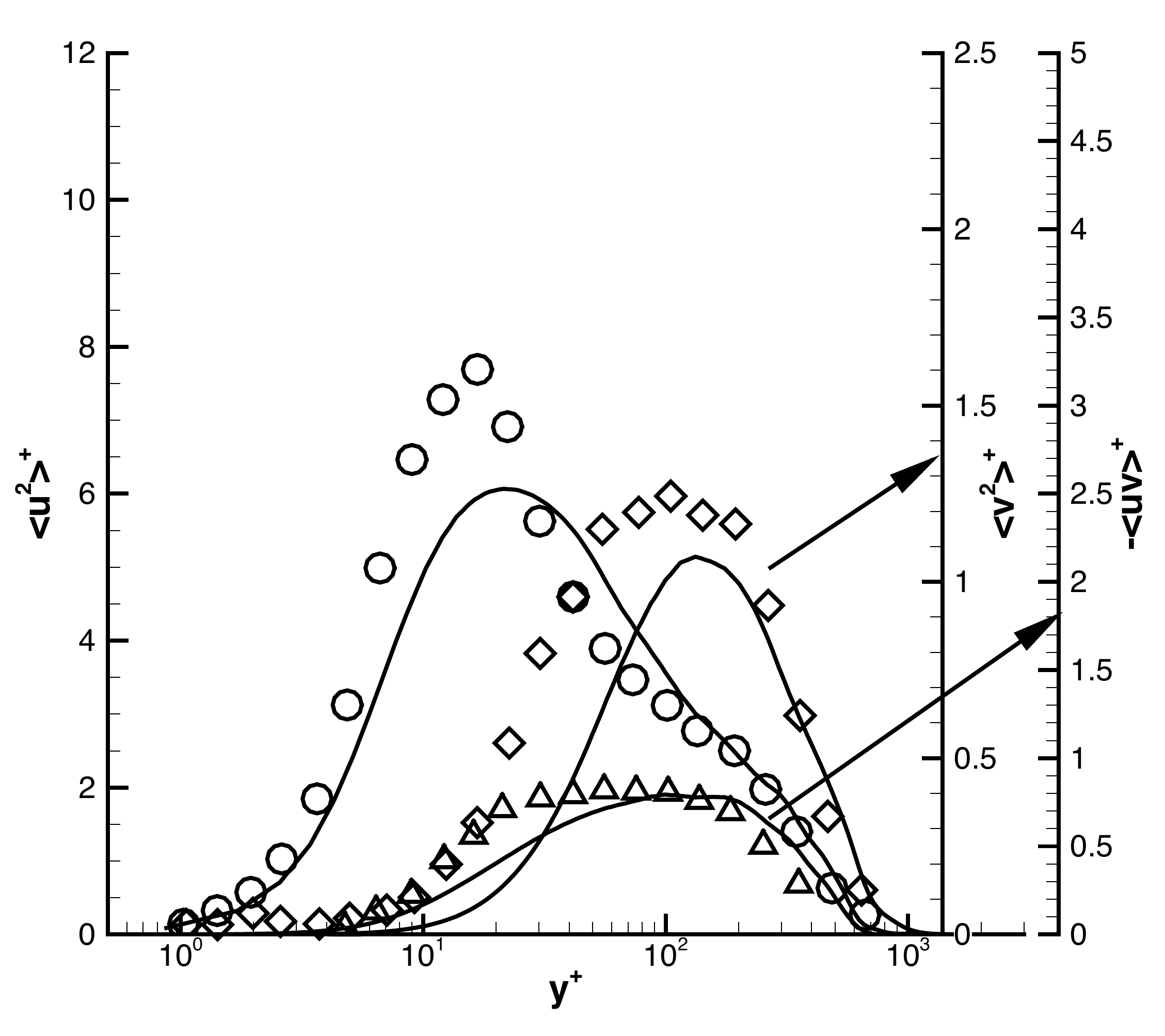}}
}
\end{minipage}
\caption{IDDES of spatially developing turbulent boundary layer: one-point statistics}
\label{IDDES_umean}
\end{figure}

The mean velocity and Reynolds stress profiles obtained with IDDES approach are shown in figure \ref{IDDES_umean}. The mean velocity profile is in excellent agreement with the DNS and experimental data. The peak in the Reynolds stresses are not predicted accurately. This is because of the near-wall RANS model used in the IDDES approach. It is a well-known issue with SST $k-\omega$ model that the near-wall peak in turbulent kinetic energy is underpredicted, but it gives accurate mean velocity \cite{durbin:2011}.

\subsubsection{Comparison of vortical structures predicted by LES and IDDES}

\begin{figure}[!htp]
\centering{
\includegraphics[width=80mm]{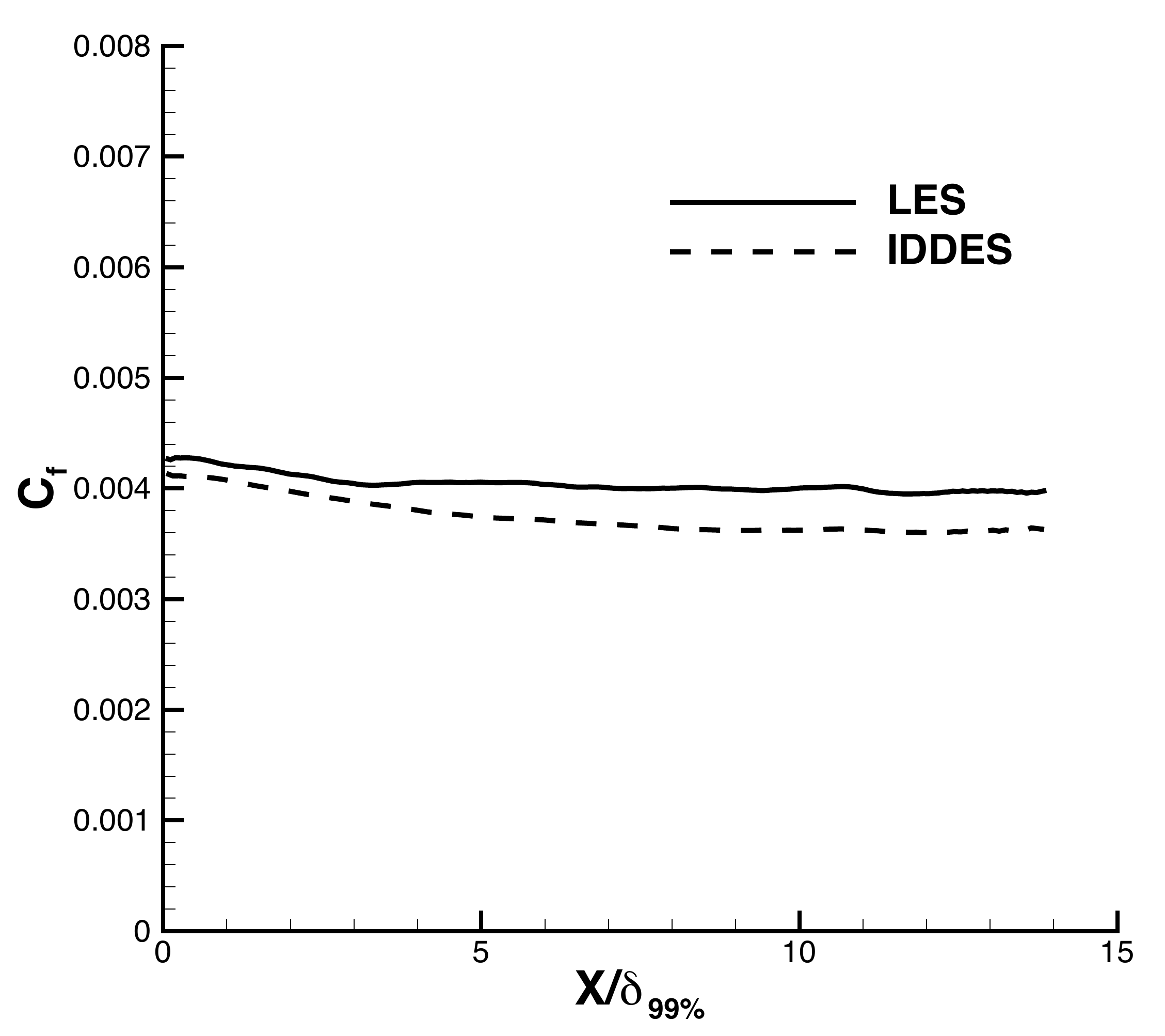}
}
\caption{Comparison of skin friction variation predicted by LES and IDDES}
\label{cf_compare}
\end{figure}

The skin friction variation along the bottom wall predicted by LES and IDDES approaches is plotted in figure \ref{cf_compare}. As the momentum thickness increases along the bottom wall, the skin friction decreases and it is predicted well by both the approaches. The quantitative discrepancy between the predicted skin friction is due to the different near-wall behavior of the underlying closure models used.

\begin{figure}[!htp]
\begin{minipage}[t!]{80mm}
\centering{
\subfigure[LES]{\includegraphics[width=80mm]{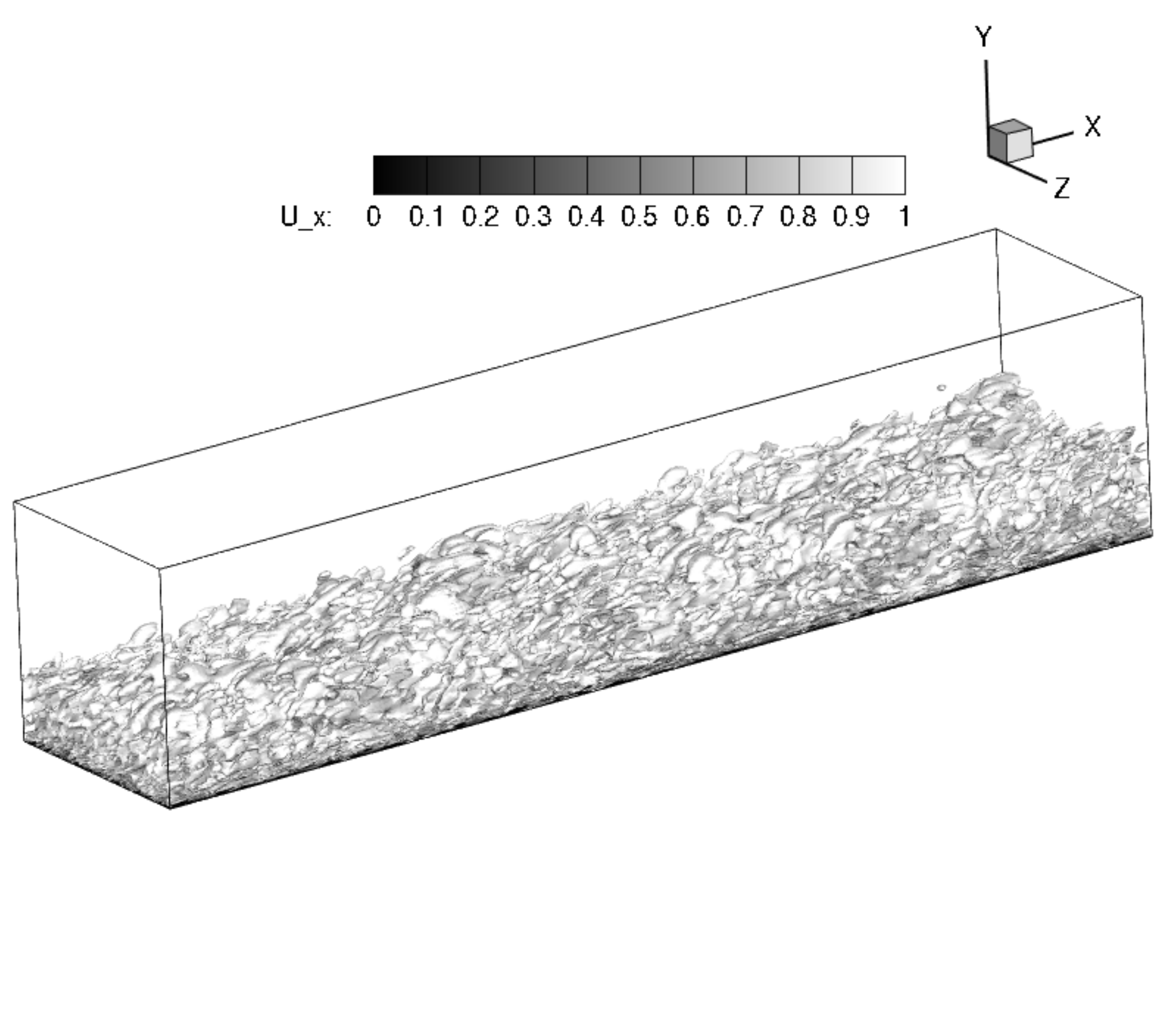}}
}
\end{minipage}
\hfill
\begin{minipage}[t!]{80mm}
\centering{
\subfigure[IDDES]{\includegraphics[width=80mm]{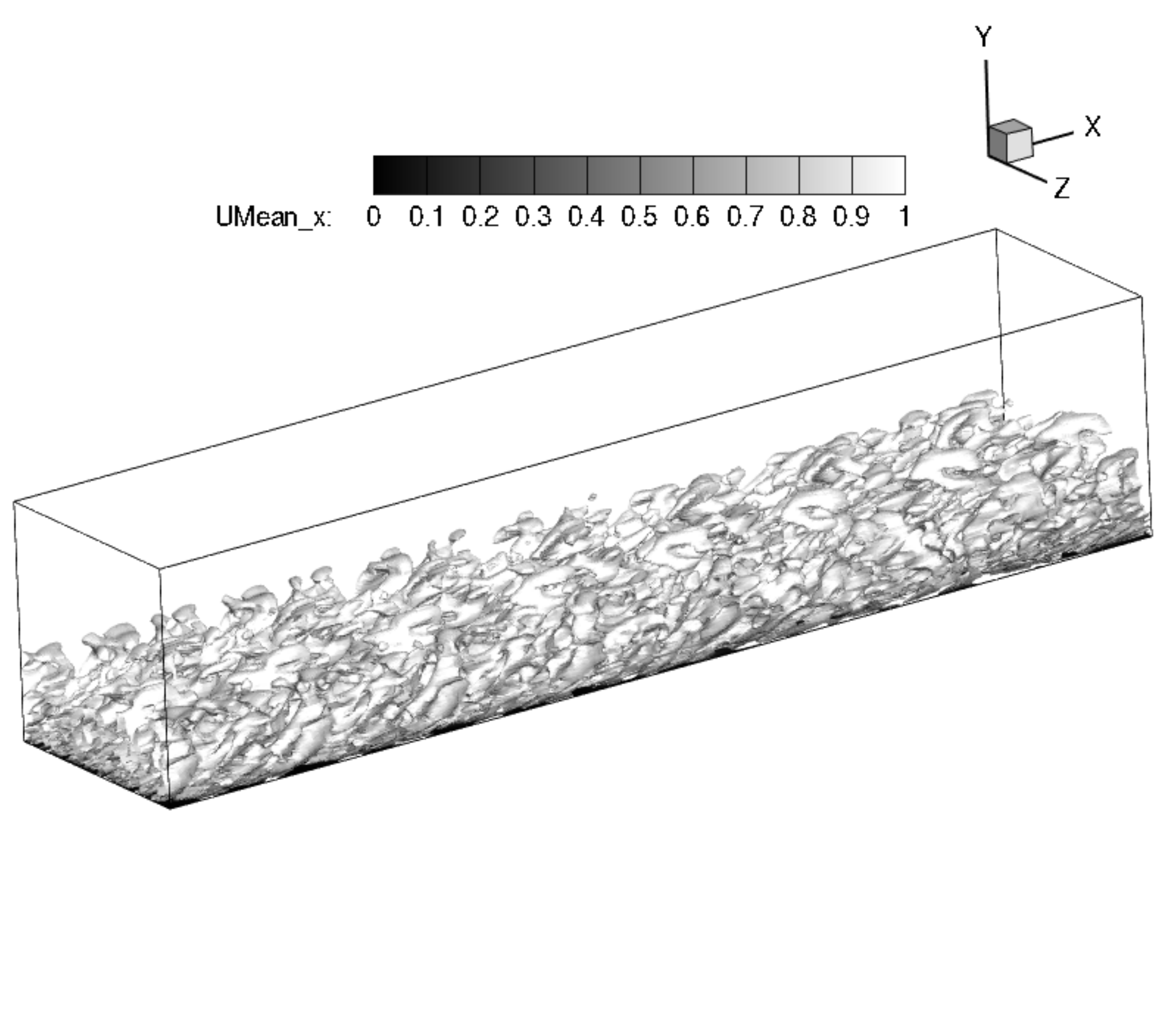}}
}
\end{minipage}
\caption{Vortical structures predicted using Q-criterion}
\label{vortical_compare}
\end{figure}

The vortical structures resolved using LES and IDDES are shown in figure \ref{vortical_compare}. As expected, LES predicts finer scale near-wall structures. In IDDES, LES is activated away from the wall and hence only large scale vortical structures are resolved. This evidence of vortical structures shows the effectiveness of the inflow generation methodology for eddy resolving simulations.

\subsection{Problem 2: LES of flow through annular diffuser}

\begin{figure}[!htp]
\begin{minipage}[t!]{80mm}
\centering{
\subfigure[Case 1: $\theta_{r}=4\%$ of $(r_{o}-r_{i})$]{\includegraphics[width=80mm]{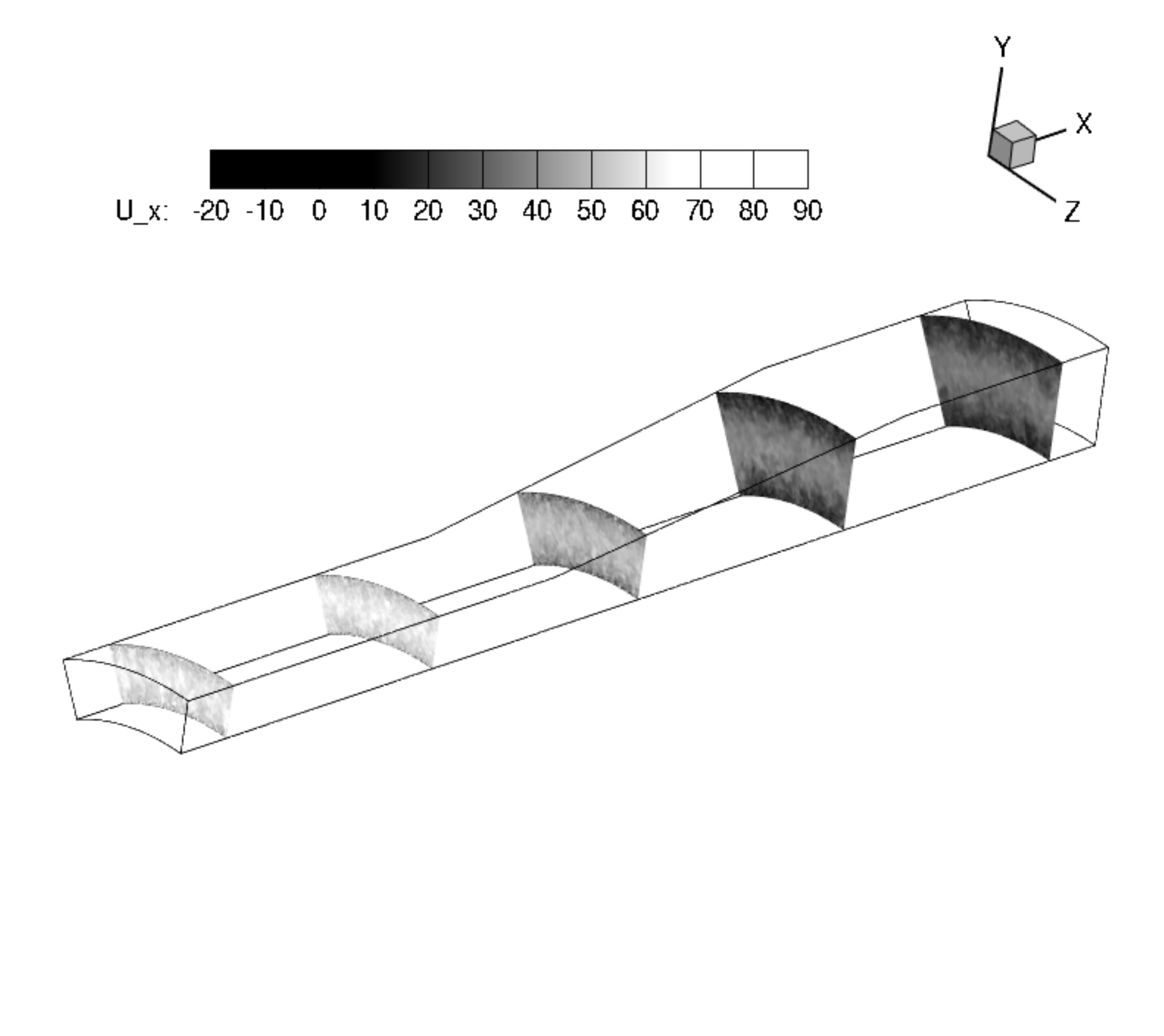}}
}
\end{minipage}
\hfill
\begin{minipage}[t!]{80mm}
\centering{
\subfigure[Case 2: $\theta_{r}=0.3\%$ of $(r_{o}-r_{i})$]{\includegraphics[width=80mm]{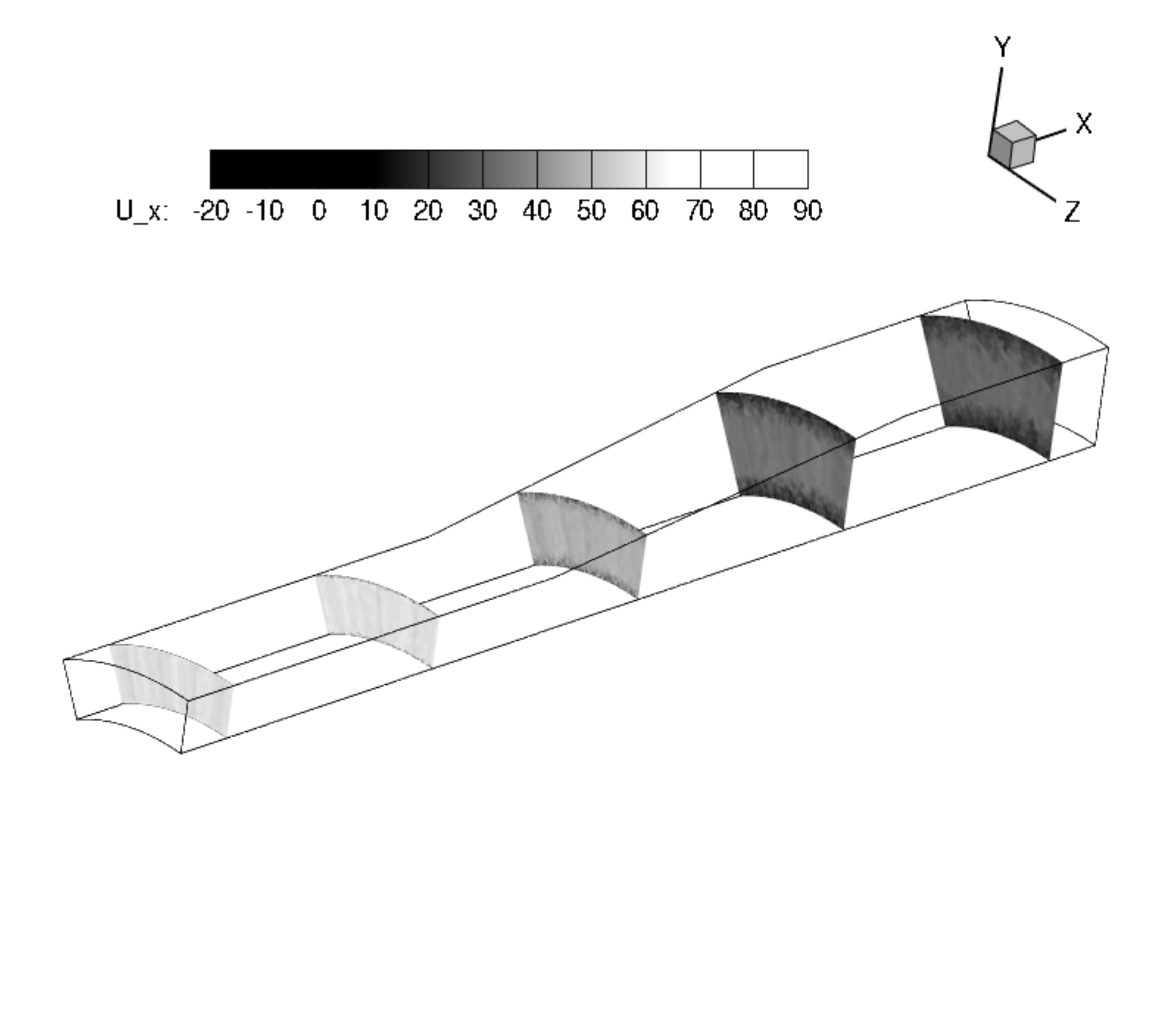}}
}
\end{minipage}
\caption{Cros plane velocity contours}
\label{contours_diffuser}
\end{figure}

\begin{figure}[!htp]
\begin{minipage}[t!]{80mm}
\centering{
\subfigure[Case 1: $\theta_{r}=4\%$ of $(r_{o}-r_{i})$]{\includegraphics[width=80mm]{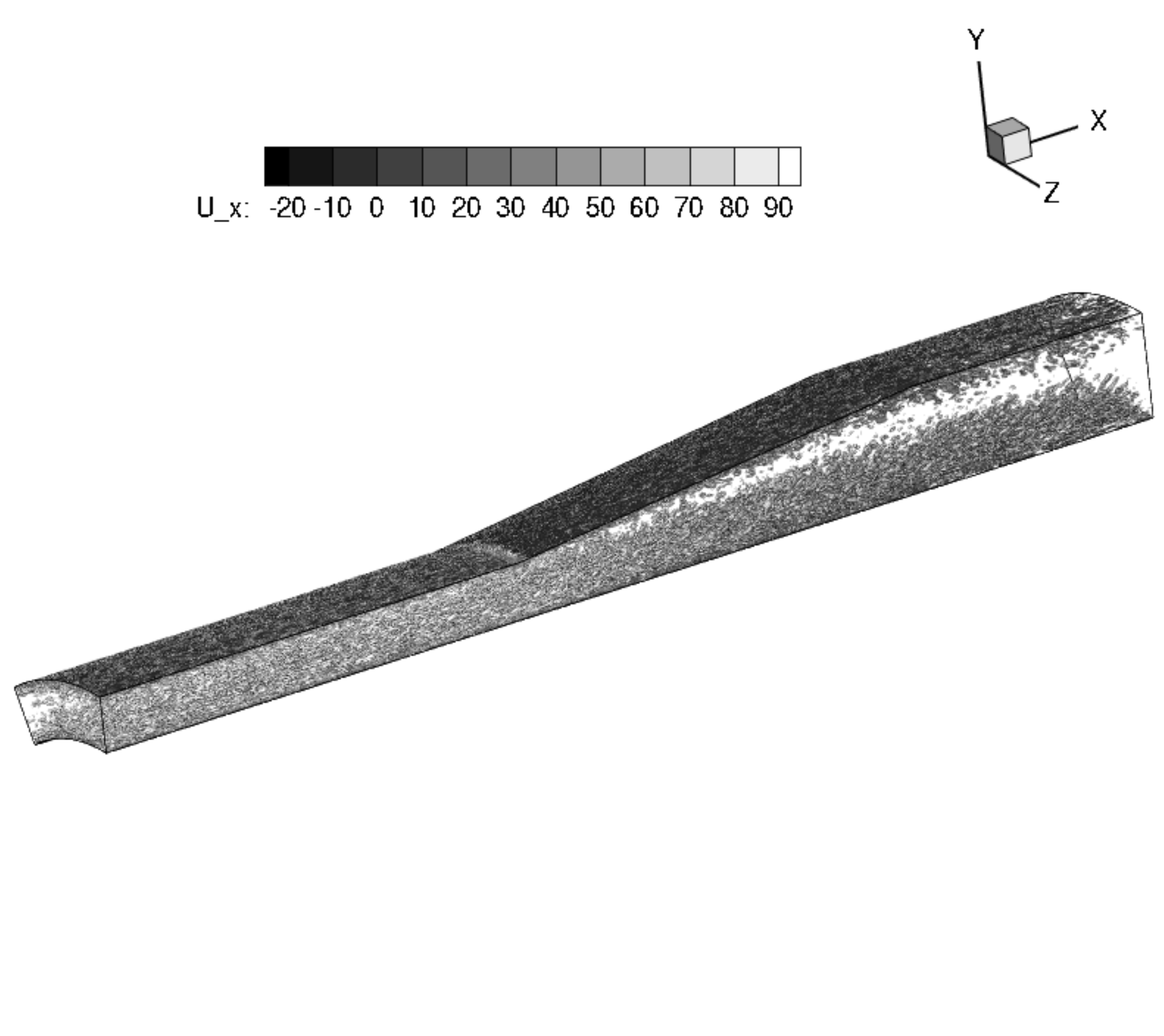}}
}
\end{minipage}
\hfill
\begin{minipage}[t!]{80mm}
\centering{
\subfigure[Case 2: $\theta_{r}=0.3\%$ of $(r_{o}-r_{i})$]{\includegraphics[width=80mm]{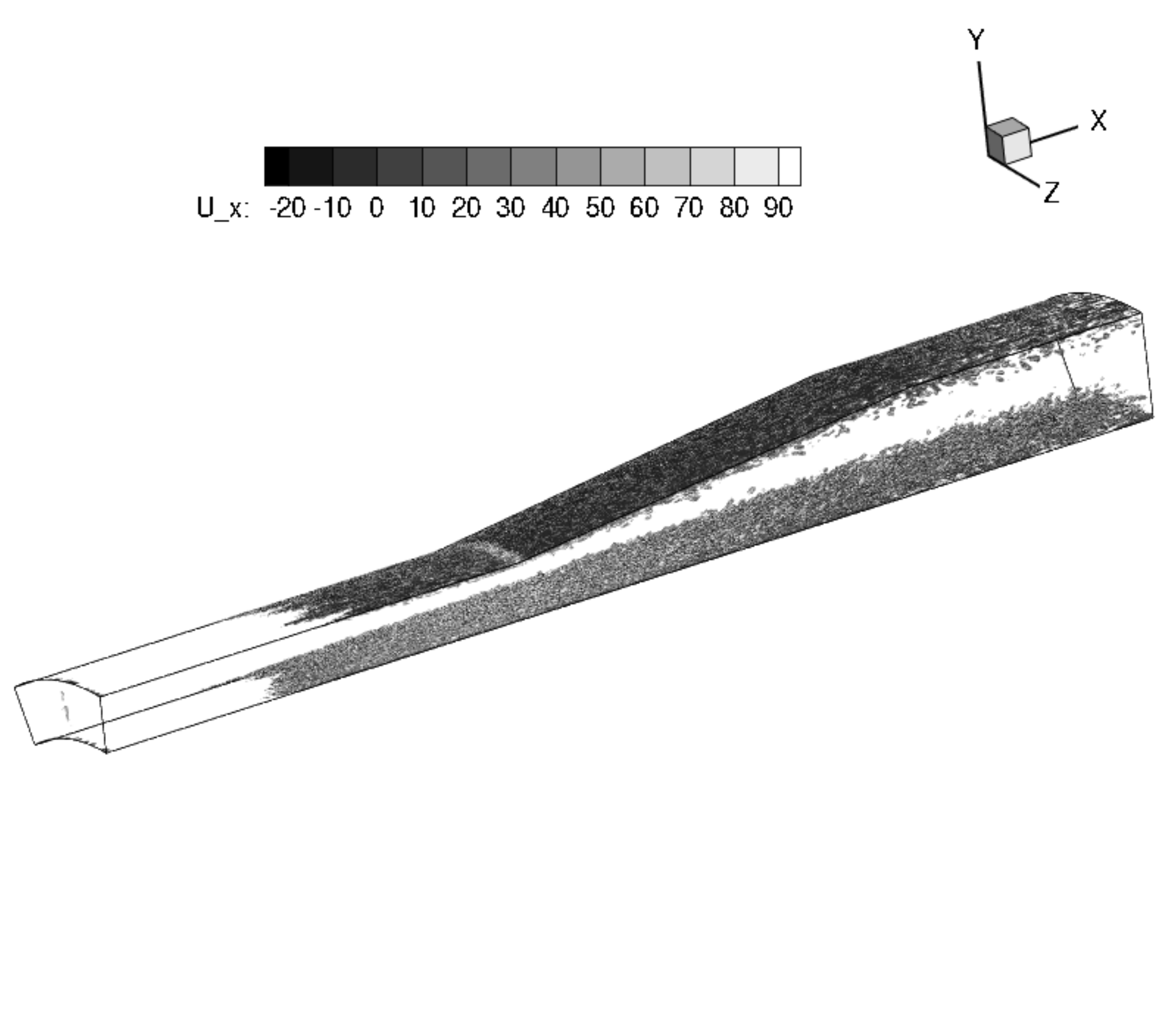}}
}
\end{minipage}
\caption{Vortical structures predicted in the annular diffuser using Q-criterion}
\label{vortical_diffuser}
\end{figure}

\begin{figure}[!htp]
\begin{minipage}[t!]{80mm}
\centering{
\subfigure[Reynolds shear stress profiles]{\includegraphics[width=80mm]{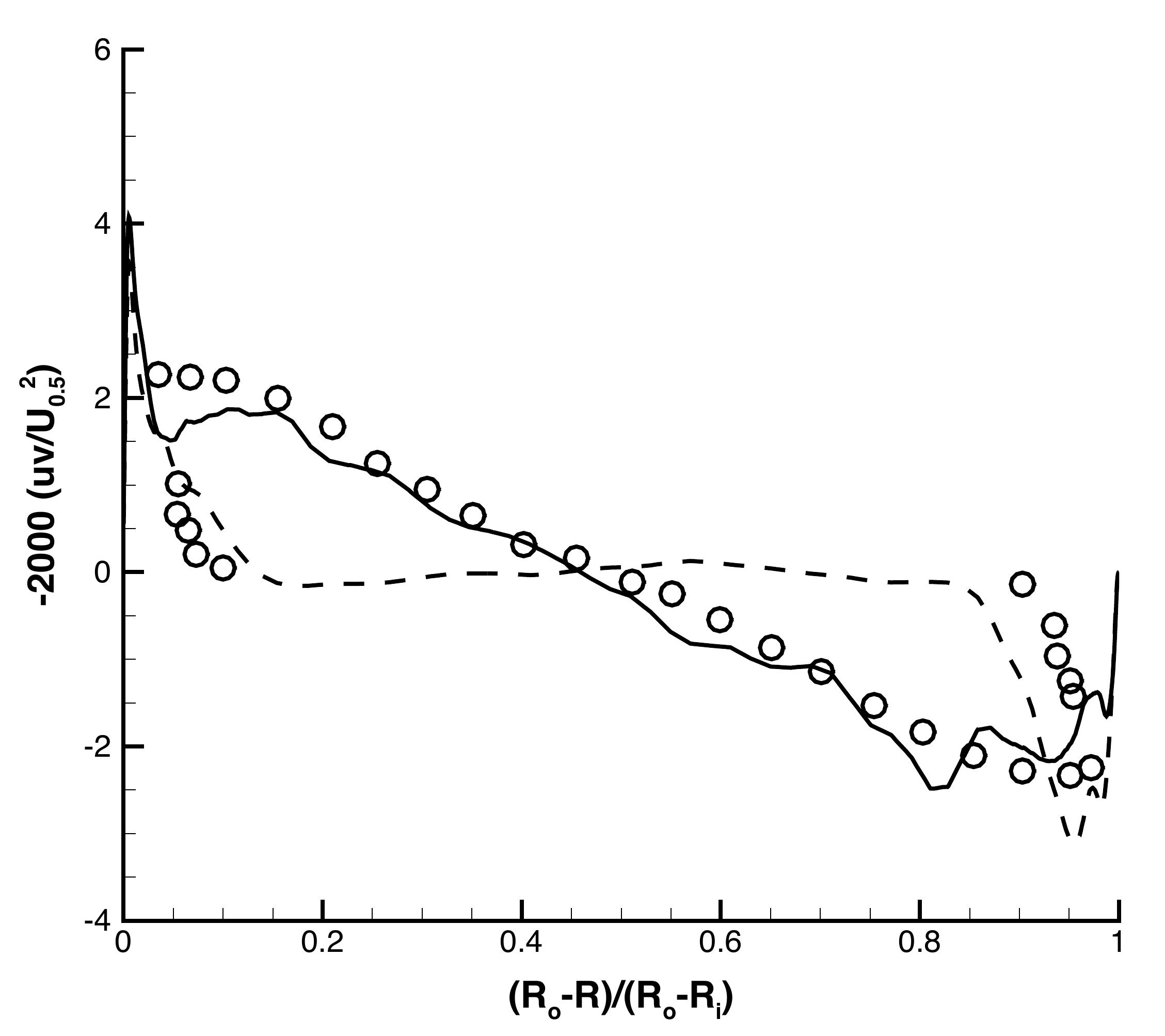}}
}
\end{minipage}
\hfill
\begin{minipage}[t!]{80mm}
\centering{
\subfigure[Mean velocity]{\includegraphics[width=80mm]{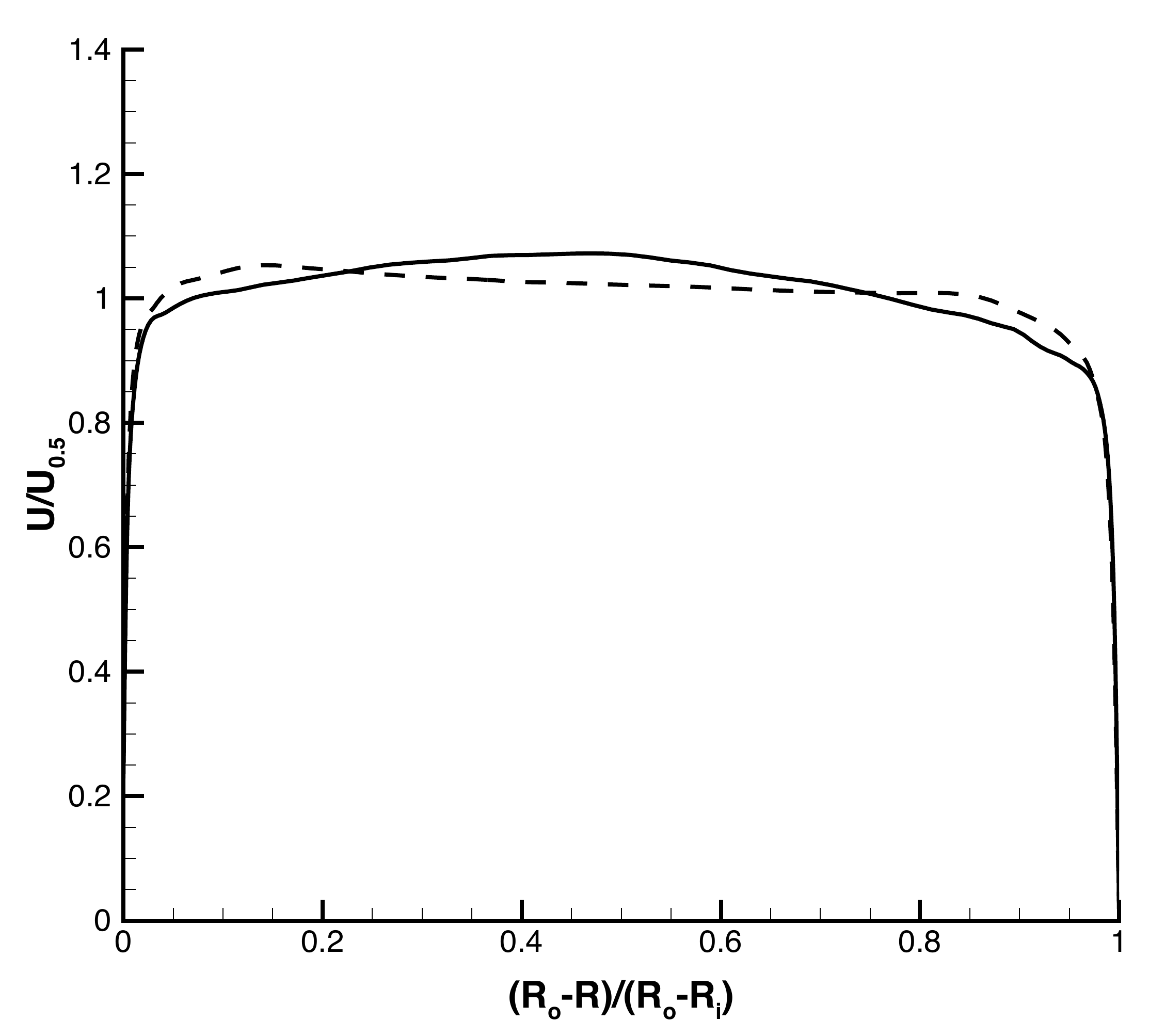}}
}
\end{minipage}
\caption{LES of $30^\circ$ sector of the annular diffuser. Profiles are extracted at the inlet to the diffuser. Solid lines: Case1, Dotted lines: Case2}
\label{umean_diffuser}
\end{figure}

To show the utility of the inflow generation algorithm for wall-bounded turbulent flows applied to turbomachinery, LES of flow through annular diffuser is performed. The geometry used is that of \cite{stevens:1980}. Two cases are considered with momentum thickness of about $4\%$ and $0.3\%$ of the height of the annulus at the inflow to the diffuser. The Reynolds shear stress profiles predicted by LES inflow generation method are compared with the available experimental data. The mesh used has $348 \times 128 \times 128$ points in the streamwise, radial, and azimuthal directions, respectively. The simulation has been run for 50 timescales ($r_{0.5}/U_{0.5}$) and the statistics are collected over another 50 inertial timescales.

The cross plane velocity contours in figure \ref{contours_diffuser} show that the growth of the boundary layer is much rapid when the inflow momentum thickness is larger. It is also evidenced from the vortical structures presented in figure \ref{vortical_diffuser}. The velocity and Reynolds shear stress profiles in these two cases are plotted in figure \ref{umean_diffuser}. The normalized shear stress shows good agreement with the experimental data at the inflow to the diffuser. The velocity profile with the lower momentum thickness looks to be uniform, but it is critical to provide the turbulence quantities to predict the flow behavior through the annular diffuser. The proposed inflow generation method proves to be effective for this purpose.

\section{Conclusions}
A simple variant of recycling and rescaling method to generate inflow turbulence is presented for unstructured grid CFD codes. This method contains a momentum thickness based rescaling algorithm combined with a mirroring method to disorganize spanwise durable structures. The mean streamwise velocity and turbulence profiles predicted by LES and IDDES proves the accuracy of the methodology. For annular diffuser, it has been demonstrated that the algorithm presented can be used to obtain required turbulent boundary layer characteristics at the inflow. It is hoped that the method presented will be useful for eddy resolving simulations of more complex practical problems.

\section{Acknowledgements}
My sincere thanks to Professor Paul Durbin for helpful discussions on this work. This work used computational resources of Stampede under the Extreme Science and Engineering Discovery Environment (XSEDE), which is supported by National Science Foundation.

\bibliographystyle{asmems4}

\bibliography{Inflow_Turbomachinery}

\begin{thebibliography}{10}

\bibitem{stevens:1980}
Stevens, S.~J., and Williams, G.~J., 1980.
\newblock ``The influence of inlet conditions on the performance of annular
  diffusers''.
\newblock {\em J. Fluids Eng., {\bf 102}}, September, pp.~357--363.

\bibitem{akselvoll:1995}
Akselvoll, K., and Moin, P., 1995.
\newblock ``Large eddy simulation of turbulent confined coannular jets and
  turbulent flow over a backward facing step''.
\newblock Report tf-63, Thermosciences Div., Dept. Mech. Eng., Stanford
  University, Stanford, CA 94305.

\bibitem{spalart:1988}
Spalart, P., 1988.
\newblock ``Direct simulation of a turbulent boundary layer up to
  ${R}_{\theta}=1410$''.
\newblock {\em J. Fluid Mech., {\bf 187}}, pp.~61--98.

\bibitem{lund:1998}
Lund, T., X., W., and Squires, K., 1998.
\newblock ``Generation of turbulent inflow data for spatially developing
  boundary layer simula- tions''.
\newblock {\em J. Comp. Phys., {\bf 140}}, pp.~233--258.

\bibitem{liu:2006}
Liu, K., and Pletcher, R.~H., 2006.
\newblock ``Inflow conditions for the large eddy simulation of turbulent
  boundary layers: a dynamic recycling procedure''.
\newblock {\em J. Comp. Phys., {\bf 219}}(1), pp.~1--6.

\bibitem{spalart:2006}
Spalart, P., Strelets, M., and Travin, A., 2006.
\newblock ``Direct numerical simulation of large-eddy-break-up devices in a
  boundary layer''.
\newblock {\em Int. J. Heat Fluid Flow, {\bf 27}}, pp.~902--910.

\bibitem{jewkes:2011}
Jewkes, J., Chung, Y., and Carpenter, P., 2011.
\newblock ``Modification to a turbulent inflow generation method for
  boundary-layer flows''.
\newblock {\em AIAA Journal, {\bf 49}}(1), pp.~247--250.

\bibitem{morgan:2011}
Morgan, B., Larsson, J., Kawai, S., and Lele, S.~K., 2011.
\newblock ``Improving low-frequency characteristics of recycling/rescaling
  inflow turbulence generation''.
\newblock {\em AIAA Journal, {\bf 49}}(3), pp.~582--587.

\bibitem{ferrante:2004}
Ferrante, A., and Elghobashi, S.~E., 2004.
\newblock ``A robust method of generating inflow conditions for direct
  simulations of spatially-developing turbulent boundary layers''.
\newblock {\em J. Comp. Phys., {\bf 198}}, pp.~372--387.

\bibitem{yao:2002}
Yao, Y.~F., and Sandham, N.~D., 2002.
\newblock ``{DNS} of turbulent flow over a bump with shock/boundary-layer
  interactions''.
\newblock In Fifth International Symposium on Engineering Turbulence Modeling
  and Measurements, pp.~677--686.

\bibitem{klein:2003}
Klein, M., Sadiki, A., and Janicka, J., 2003.
\newblock ``A digital filter based generation of inflow data for spatially
  developing direct numerical or large eddy simulations''.
\newblock {\em J. Comp. Phys., {\bf 186}}(2), pp.~652--665.

\bibitem{jarrin:2006}
Jarrin, N., Benhamadouche, S., Laurence, D., and Prosser, R., 2006.
\newblock ``A synthetic-eddy-method for generating inflow conditions for large
  eddy simulations''.
\newblock {\em Int. J. Heat Fluid Flow, {\bf 27}}(4), pp.~585--593.

\bibitem{pamies:2009}
Pami\'{e}s, M., P., W., Gamier, E., Dick, S., and Sagaut, P., 2009.
\newblock ``Generation of synthetic turbulent inflow data for large eddy
  simulation of spatially evolving wall-bounded flows''.
\newblock {\em Phys. Fluids, {\bf 21}}(4), pp.~045103--1--045103--15.

\bibitem{keating:2004}
Keating, A., Piomelli, U., Balaras, E., and Kaltenbach, H., 2004.
\newblock ``\textit{A priori} and \textit{a posteriori} tests of inflow
  conditions for large-eddy simulation''.
\newblock {\em Phys. Fluids, {\bf 16}}, p.~4696.

\bibitem{spalart:2006a}
Spalart, P., Deck, S., Shur, M.~L., Squires, K.~D., Strelets, M.~K., and
  Travin, A., 2006.
\newblock ``A new version of detached-eddy simulation, resistant to ambiguous
  grid densities''.
\newblock {\em Theor. Comput. Fluid Dyn., {\bf 20}}, pp.~181--195.

\bibitem{arolla:2014}
Arolla, S.~K., and Durbin, P.~A., 2014.
\newblock ``Generating inflow turbulence for eddy simulation of turbomachinery
  flows''.
\newblock In 52nd AIAA Aerospace Sciences Meeting, National Harbor, Maryland.

\bibitem{gritskevich:2012}
Gritskevich, M.~S., Garbaruk, A.~V., Sch\"{u}tze, J., and Menter, F.~R., 2012.
\newblock ``Development of {DDES} and {IDDES} formulations of the k-$\omega$
  shear stress transport model''.
\newblock {\em Flow Turbul. Combust., {\bf 88}}, pp.~431--449.

\bibitem{germano:1991}
Germano, M., Piomelli, U., Moin, P., and Cabot, W.~H., 1991.
\newblock ``A dynamic subgrid-scale eddy viscosity model''.
\newblock {\em Phys. Fluids A, {\bf 3}}(7), p.~1760.

\bibitem{lilly:1992}
Lilly, D.~K., 1992.
\newblock ``A proposed modification of the germano subgrid-scale closure
  method''.
\newblock {\em Phys. Fluids A, {\bf 4}}, p.~633.

\bibitem{degraaff:2000}
De{G}raaff, D.~B., and Eaton, J.~K., 2000.
\newblock ``Reynolds-number scaling of the flat-plate turbulent boundary
  layer''.
\newblock {\em J. Fluid Mech., {\bf 422}}, pp.~319--346.

\bibitem{durbin:2011}
Durbin, P.~A., and Pettersson~Reif, B.~A., 2011.
\newblock {\em Statistical theory and mathematical modeling for turbulent
  flows}.
\newblock John Wiley $\&$ Sons.

\end{thebibliography}

\end{document}